\documentclass[twocolumn]{aastex62}

\usepackage{graphicx}
\usepackage{amssymb}
\usepackage{amsmath}
\usepackage{natbib}
\usepackage{wrapfig}
\usepackage[utf8]{inputenc}
\usepackage[T1]{fontenc}
\usepackage{url}
\newcommand{\masyr}{ \ {\rm{mas \ yr^{-1}}}\>}
\newcommand{\kms}{ \ {\rm{km \ s^{-1}}}\>}

\accepted{to ApJ September 4, 2018}

\shorttitle{LMC Companions}
\shortauthors{Kallivayalil et al.}

\begin{document}

\title{The Missing Satellites of the Magellanic Clouds? Gaia Proper Motions of the Recently Discovered Ultra-Faint Galaxies}

\correspondingauthor{Nitya Kallivayalil}
\email{njk3r@virginia.edu}

\author[0000-0002-3204-1742]{Nitya Kallivayalil}
\affiliation{Department of Astronomy, University of Virginia, 530 McCormick Road, Charlottesville, VA 22904, USA}

\author{Laura V. Sales}
\affiliation{Department of Physics and Astronomy, University of California Riverside, 900 University Avenue, CA 92507, USA}

\author{Paul Zivick}
\affiliation{Department of Astronomy, University of Virginia, 530 McCormick Road, Charlottesville, VA 22904, USA}

\author{Tobias K. Fritz}
\affiliation{Instituto de Astrofisica de Canarias, Calle Via Lactea s/n, E-38205 La Laguna, Tenerife, Spain}
\affiliation{Universidad de La Laguna, Dpto. Astrofisica, E-38206 La Laguna, Tenerife, Spain}

\author{Andr\'es Del Pino}
\affiliation{Space Telescope Science Institute, 3700 San Martin Drive, Baltimore, MD 21218, USA}

\author{Sangmo Tony Sohn}
\affiliation{Space Telescope Science Institute, 3700 San Martin Drive, Baltimore, MD 21218, USA}

\author{Gurtina Besla}
\affiliation{Steward Observatory, University of Arizona, 933 North Cherry Avenue, Tucson, AZ 85721, USA}

\author{Roeland P. van der Marel}
\affiliation{Space Telescope Science Institute, 3700 San Martin Drive, Baltimore, MD 21218, USA}
\affiliation{Center for Astrophysical Sciences, Department of Physics \& Astronomy, Johns Hopkins University, Baltimore MD 21218, USA}

\author{Julio F. Navarro}
\affiliation{Senior CIfAR Fellow, Department of Physics and Astronomy, University of Victoria, Victoria, BC V8P 5C2, Canada}

\author{Elena Sacchi}
\affiliation{Space Telescope Science Institute, 3700 San Martin Drive, Baltimore, MD 21218, USA} 

\begin{abstract}

According to LCDM theory, hierarchical evolution occurs on all mass
scales, implying that satellites of the Milky Way should also have
companions. The recent discovery of ultra-faint dwarf galaxy
candidates in close proximity to the Magellanic Clouds provides an
opportunity to test this theory. We present proper motion (PM) measurements
for 13 of the 32 new dwarf galaxy candidates using Gaia data release
2. All 13 also have radial velocity measurements. We compare the
measured 3D velocities of these dwarfs to those expected at the
corresponding distance and location for the debris of an LMC analog in
a cosmological numerical simulation. We conclude that 4 of these
galaxies (Hor1, Car2, Car3 and Hyi1) have come in with the Magellanic
Clouds, constituting the first confirmation of the type of
satellite infall predicted by LCDM. Ret2, Tuc2 and Gru1 have 
velocity components that are not consistent within 3 sigma of our
predictions and are therefore less favorable. Hya2 and Dra2 could be
associated with the LMC and merit further attention. We rule out Tuc3,
Cra2, Tri2 and Aqu2 as potential members. Of the dwarfs without
measured PMs, 5 of them are deemed unlikely on the basis of their
positions and distances alone as being too far from the orbital
plane expected for LMC debris (Eri2, Ind2, Cet2, Cet3 and
Vir1). For the remaining sample, we use the simulation to predict PMs and radial velocities, finding that Phx2 has an
overdensity of stars in DR2 consistent with this PM prediction.

\end{abstract}


\section{Introduction} \label{sec:intro}

In the prevailing $\Lambda$Cold Dark Matter cosmology (LCDM), dark matter (DM) halos build up their mass by the accretion of smaller objects. The satellites that orbit our Milky Way (MW) are remnants of this process, giving us insights into the accretion events that built our Galaxy’s DM halo. Self-similarity dictates that this hierarchical evolution should occur on all mass scales, implying that satellites of the MW also once had companions. However, testing this picture is challenging -- faint companions of dwarf galaxies are difficult to detect at large distances and the tidal field of the MW acts to disrupt such groups over multiple orbits. The recent discoveries of 32 candidate low mass, dwarf galaxies in close proximity on the sky to the Magellanic Clouds (\citealt{Bechtol2015}; \citealt{Drlica2015}; \citealt{koposov15}; \citealt{Martin2015}; \citealt{Laevens2015}; \citealt{Torrealba2016a}; \citealt{torrealba16b}; \citealt{Torrealba2018}; \citealt{Kim2015a}; \citealt{Kim2015b}; \citealt{drlica-wagner16}; \citealt{Koposov2018}; \citealt{Homma2018}) present a new opportunity to test this theory.

Our understanding of the orbital history of the Magellanic Clouds has also gone through a recent revision. The Clouds are moving faster with respect to the MW than previously believed \citep{NK06b, NK06a, NK13}, implying that, instead of being long term companions, they are likely on their first infall towards the MW \citep{besla07}. Such late infall is not unexpected within the LCDM paradigm \citep{boylankolchin11}. This scenario further implies that MW tides have not had sufficient time to disrupt the infalling system, and so companions of the Clouds should share common orbital properties \citep[][hereafter S17]{sales11,sales17}.


A massive satellite such as the LMC is expected to host its own satellites at infall \citep[e.g.,][]{Lynden-Bell1976,Li2008b,donghia2008}. Using semi-analytic models of galaxy formation \citet{Sales2013} determined the number of satellites expected around dwarf galaxies like the Large Magellanic Cloud (LMC) in LCDM theory (see also \citealt{guo10,Guo2011}). The LMC has a stellar mass of $M_\star \sim 3 \times 10^9 M_\odot$ \citep{vdM02}, implying a DM mass of log($M_{ \rm dark}$) = 10.75--11.25 prior to capture by the MW \citep{moster13,NK13}. From \citet{Sales2013}, such LMC analogs should host one Small Magellanic Cloud (SMC) analog (mass ratio $\sim 0.1$) in addition to $\sim 5$--40 satellites that are $\times 0.001$ its mass, i.e., ultra-faint dwarf galaxies: $M_\star \sim 0.1 - 1 \times 10^4 M_\odot$, log($M_{ \rm dark}) \sim 8$. This begs the question: \textit{where are these galaxies today?}

\citet{Wetzel2015} used the ELVIS suite of cosmological simulations of MW analogs \citep{Garrison-Kimmel2014} to trace the orbital histories of surviving satellites at $z = 0$. They determined that half of the current low mass ($M_\star < 10^6 M_\odot$) satellites of the MW were preprocessed in a group prior to capture by the MW. The identification of the surviving satellite population of the LMC provides a crucial testing ground for such theories, and for the halo occupation function at low mass scales.

The discovery of dwarf galaxies in the vicinity of the Clouds could help answer these questions. The newly-discovered dwarfs range in visual magnitudes from $M_V = -2$ to $-8$, and have half-light radii $r_h \sim 20 - 1100$ pc. While some of these candidates have yet to be confirmed as DM-dominated galaxies using velocity dispersion measurements, they generally have magnitudes and sizes consistent with those of known ultra faint dwarf galaxies orbiting the MW and M31 \citep{McConnachie2012}. 

A few past works have attempted to assess membership of these candidates to the Clouds. \citet{Deason2015} used a statistical argument based on abundance-matching models applied to massive subhalos in the ELVIS simulations to suggest that 2--4 of the 9 then known candidates might have come into the MW with the LMC. \citet{yozin15} on the other hand, conclude, on the basis of orbit models, that the majority of those dwarfs could have been at least loosely associated with the Clouds. \citet{jethwa16a} constructed a dynamical model for the Magellanic Clouds satellite population based on numerical simulations. They compared this to the observed 3D spatial distribution of the candidate dwarfs, excluding likely globular clusters. They inferred that at $1\sigma$-confidence, 50\% of the candidates have $>$ 70\% probability of a Magellanic Clouds association. \citet{Dooley2017} used the abundance matching method to determine what fraction of the new ultra-faint dwarfs could be associated to the LMC depending on the assumed relation between stellar mass and halo mass. 

In a different approach, S17 (using an extension of the set up in \citealt{sales11}) identified an LMC analog in a fully cosmological simulation of a MW-sized halo in LCDM. They tracked the positions and velocities of subhalo particles to constrain the likely location in phase space of systems with prior association with the LMC, finding that of the 6 systems with kinematic data at the time, only Hydra II and Hor 1 had distances and radial velocities consistent with a Magellanic origin. Of the remaining dwarfs, six (Hor 2, Eri 3, Ret 3, Tuc 4, Tuc 5, and Phx 2) had positions and distances consistent with a Magellanic origin, but kinematic data were needed to substantiate that possibility. Conclusive evidence for association would require proper motions to constrain the orbital angular momentum direction, which, for true Magellanic satellites, must coincide with that of the Clouds. 

Here, we measure proper motions (PMs) from data release 2 (DR2) of the \textit{Gaia} survey \citep{Brown2018} for 13 of the 32 newly discovered dwarfs. We combine these proper motions with the measured radial velocities from other studies, and compare to the predictions from the \citet{sales17} set up to confirm or rule out the dynamical association of these dwarfs with the Magellanic system. 

In Section~\ref{sec:data} we describe our methodology to select member stars for each of the dwarfs and to measure their PMs. In Section~\ref{sec:results} we compare the resultant Galactocentric velocities to that of LMC debris at the positions and distances of these dwarf galaxies in the simulation. 
 We discuss and conclude in Section~\ref{sec:conc}.

\begin{table*}[t]
\begin{center}
\caption{Dwarf Galaxy Properties.\label{tbl-2}}
\begin{tabular}{lcrrrrrcrccc}
\tableline
 \\
ID & name & $l$ (deg) & $b$ (deg) & ra (deg) & dec (deg) & D (kpc) & $m - M$ & M$_V$ & r$_h$ ($\arcmin$)  & RV (km s$^{-1}$) & notes \\ 

\tableline

1 & Dra2 & 98.3 & 42.9 & 238.2 & 64.6 & 20.0 & 16.9 $\pm$ 0.3 & $-$2.9 & 2.7 & $-$347.6 $\pm$ 1.8 & [1];RV:[2] \\ 
2 & Tuc3 & 315.4 & $-$56.2 & 359.1 & $-$59.6 & 25.0 & 17.0 $\pm$ 0.2 & $-$2.4 & 6.0 & $-$102.3 $\pm$ 2.0 & [3];RV:[4] \\ 
3 & Hyi1 & 304.5 & $-$16.5 & 37.4 & $-$79.3 & 27.6 & 17.2 $\pm$ 0.0 & $-$4.7 & 7.4 & 80.4 $\pm$ 0.6 & [5] \\ 
4 & Car3 & 270.0 & $-$16.8 & 114.6 & $-$57.9 & 27.8 & 17.2 $\pm$ 0.1 & $-$2.4 & 3.8 & 284.6 $\pm$ 3.4 & [6];RV:[7] \\ 
5 & Ret2 & 265.9 & $-$49.6 & 53.9 & $-$54.0 & 30.0 & 17.4 $\pm$ 0.2 & $-$2.7 & 3.6 & 62.8 $\pm$ 0.5 & [8];RV:[9] \\ 
6 & Cet2 & 156.5 & $-$78.5 & 19.5 & $-$17.4 & 30.0 & 17.4 $\pm$ 0.2 & 0.0 & 1.9 & - & [3];RV:[10] \\ 
7 & Tri2 & 140.9 & $-$23.8 & 33.3 & 36.2 & 30.0 & 17.4 $\pm$ 0.1 & $-$1.8 & 3.9 & $-$381.7 $\pm$ 1.1 & [11];RV:[12] \\ 
8 & Car2 & 270.0 & $-$17.1 & 114.1 & $-$58.0 & 36.2 & 17.8 $\pm$ 0.1 & $-$4.5 & 8.7 & 477.2 $\pm$ 1.2 & [6];RV:[7] \\ 
9 & Pic2 & 269.6 & $-$24.1 & 101.2 & $-$59.9 & 45.0 & 18.3 $\pm$ 0.1 & $-$3.2 & 3.6 & - & [13] \\ 
10 & Tuc4 & 313.3 & $-$55.3 & 0.7 & $-$60.9 & 48.0 & 18.4 $\pm$ 0.2 & $-$3.5 & 9.1 & - & [3] \\ 
11 & Gru2 & 351.1 & $-$51.9 & 331.0 & $-$46.4 & 53.0 & 18.6 $\pm$ 0.2 & $-$3.9 & 6.0 & - & [3] \\ 
12 & Tuc5 & 316.3 & $-$51.9 & 354.3 & $-$63.3 & 55.0 & 18.7 $\pm$ 0.3 & $-$1.6 & 1.0 & - & [3;26] \\ 
13 & Tuc2 & 327.9 & $-$52.8 & 343.0 & $-$58.6 & 57.0 & 18.8 $\pm$ 0.2 & $-$3.8 & 9.8 & $-$129.1 $\pm$ 3.5 & [8];RV:[14] \\ 
14 & Sag2 & 18.9 & $-$22.9 & 298.2 & $-$22.1 & 67.0 & 19.1 $\pm$ 0.1 & $-$5.2 & 2.0 & - & [11] \\ 
15 & Lae3 & 63.6 & $-$21.2 & 316.8 & 15.0 & 67.0 & 19.1 $\pm$ 0.1 & $-$4.4 & 0.4 & - & [1] \\ 
16 & Hor2 & 262.5 & $-$54.1 & 49.1 & $-$50.0 & 78.0 & 19.5 $\pm$ 0.2 & $-$2.6 & 2.1 & - & [15] \\ 
17 & Hor1 & 270.9 & $-$54.7 & 43.9 & $-$54.1 & 79.0 & 19.5 $\pm$ 0.2 & $-$3.4 & 1.3 & 112.8 $\pm$ 2.6 & [8];RV:[16] \\ 
18 & Phx2 & 323.3 & $-$60.2 & 355.0 & $-$54.4 & 83.0 & 19.6 $\pm$ 0.2 & $-$2.8 & 1.1 & - & [8] \\ 
19 & Eri3 & 274.3 & $-$59.6 & 35.7 & $-$52.3 & 87.0 & 19.7 $\pm$ 0.2 & $-$2.0 & 0.5 & - & [8;26] \\ 
20 & Vir1 & 276.9 & 59.6 & 180.0 & $-$0.7 & 91.0 & 19.8 $\pm$ 0.2 & $-$0.3 & 1.8 & - & [17] \\ 
21 & Ret3 & 273.9 & $-$45.6 & 56.4 & $-$60.4 & 92.0 & 19.8 $\pm$ 0.3 & $-$3.3 & 2.4 & - & [3] \\ 
22 & Ind1 & 347.3 & $-$42.6 & 317.2 & $-$51.2 & 100.0 & 20.0 $\pm$ 0.2 & $-$3.5 & 1.3 & - & [8] \\ 
23 & Aqu2 & 55.1 & $-$53.0 & 338.5 & $-$9.3 & 107.9 & 20.2 $\pm$ 0.1 & $-$4.4 & 5.1 & $-$71.1 & [18];RV:[18] \\ 
24 & Pic1 & 257.1 & $-$40.4 & 70.9 & $-$50.3 & 114.0 & 20.3 $\pm$ 0.2 & $-$3.1 & 0.9 & - & [8] \\ 
25 & Cra2 & 282.9 & 42.0 & 177.3 & $-$18.4 & 117.5 & 20.4 $\pm$ 0.0 & $-$8.2 & 31.2 & 87.5 $\pm$ 0.4 & [19];RV:[20] \\ 
26 & Gru1 & 338.6 & $-$58.8 & 344.2 & $-$50.2 & 120.0 & 20.4 $\pm$ 0.2 & $-$3.4 & 1.8 & $-$140.5 $\pm$ 2.4 & [8];RV:[14] \\ 
27 & Hya2 & 295.6 & 30.5 & 185.4 & $-$32.0 & 134.0 & 20.6 $\pm$ 0.2 & $-$4.8 & 1.7 & 303.1 $\pm$ 1.4 & [21];RV:[22] \\ 
28 & Col1 & 231.6 & $-$28.9 & 82.9 & $-$28.0 & 182.0 & 21.3 $\pm$ 0.2 & $-$4.5 & 1.9 & - & [3] \\ 
29 & Ind2 & 354.0 & $-$37.4 & 309.7 & $-$46.2 & 214.0 & 21.6 $\pm$ 0.2 & $-$4.3 & 2.9 & - & [3] \\ 
30 & Peg3 & 69.8 & $-$41.8 & 336.1 & 5.4 & 215.0 & 21.7 $\pm$ 0.1 & $-$3.4 & 0.8 & $-$222.9 $\pm$ 2.6 & [23];RV:[24] \\ 
31 & Cet3 & 163.8 & $-$61.1 & 31.3 & $-$4.3 & 251.0 & 22.0 $\pm$ 0.2 & $-$2.5 & 1.2 & - & [17] \\ 
32 & Eri2 & 249.4 & $-$51.4 & 56.1 & $-$43.5 & 380.0 & 22.9 $\pm$ 0.2 & $-$6.6 & 1.5 & 75.6 $\pm$ 2.4 & [8];RV:[25] \\

 \tableline
\end{tabular}
\tablecomments{Properties of the Dwarf galaxies that are the subject of this study listed in order of increasing distance. Column 1 lists an ID, followed by our naming convention, galactic longitude and latitude ($l$, $b$), RA and DEC, heliocentric distance, distance modulus, absolute magnitude, half-light radius, and radial velocity (if measured). Citations are given in the notes, and refer to: [1] \citealt{Laevens2015}, 
[2] \citealt{Martin2016}, 
[3] \citealt{Drlica2015}, 
[4] \citealt{Simon2017}, 
[5] \citealt{Koposov2018}, 
[6] \citealt{Torrealba2018}, 
[7] \citealt{Li2017}, 
[8] \citealt{Koposov2015a}, 
[9] \citealt{simon15}, 
[10] \citealt{Conn2018b}, 
[11] \citealt{Laevens2015a}, 
[12] \citealt{Kirby2017b}, 
[13] \citealt{drlica-wagner16}, 
[14] \citealt{walker16}, 
[15] \citealt{Kim2015a}, 
[16] \citealt{Koposov2015b}, 
[17] \citealt{Homma2018},
[18] \citealt{torrealba16b}, 
[19] \citealt{Torrealba2016a}, 
[20] \citealt{Caldwell2017}, 
[21] \citealt{Martin2015}, 
[22] \citealt{Kirby2015b}, 
[23] \citealt{Kim2015b}, 
[24] \citealt{Kim2016b}, 
[25] \citealt{Li2018},
[26] \citealt{Conn2018}.}
\label{tab:sample}
\end{center}
\end{table*}

\begin{table*}[t]
\begin{center}
\caption{PM Measurements.\label{pmmeas}}
\begin{tabular}{ccccccc}
\tableline
 \\

\vspace{0.1cm} Name  &   Nspec & Nnew	& $\mu_{\alpha^{\ast}}$  &  $\delta_{\mu \alpha^{\ast}}$  &   $\mu_{\delta}$   &  $\delta_{\mu \delta}$ \\

\vspace{0.1cm}  &    & 	& (mas yr$^{-1}$)  &  (mas yr$^{-1}$)  &   (mas yr$^{-1}$)   &  (mas yr$^{-1}$) \\

\tableline

\vspace{0.1cm}  Dra2   &  6 & 4   &  1.165     &      0.260     &      0.866     &      0.270  \\
\vspace{0.1cm}  Tuc3   &  20 & 12    	&     $-$0.026    &       0.037    &      $-$1.679    &       0.039  \\
\vspace{0.1cm}  Hyi1   &   26 & 8       &    3.773     &     0.032      &    $-$1.581      &     0.030  \\
\vspace{0.1cm}  Car3   &   3 & 1        &     3.065     &      0.095     &      1.567     &      0.104  \\
\vspace{0.1cm}  Ret2   &  22 & 3        &     2.398     &      0.039     &     $-$1.319     &      0.048  \\
\vspace{0.1cm}  Tri2   &  3 & 0         &     0.588     &      0.187     &      0.554     &      0.161  \\
\vspace{0.1cm}  Car2   &  15 & 14       &     1.802     &      0.038     &      0.084     &      0.038  \\
\vspace{0.1cm}  Tuc2   &  16 & 8        &     0.966     &      0.049     &     $-$1.380     &      0.062  \\
\vspace{0.1cm}  Hor1   &   4 & 5	&     0.926    &      0.070     &     $-$0.569     &     0.065  \\
\vspace{0.1cm}  Aqu2   &   2 & 3        &    $-$0.491    &      0.306     &     $-$0.049     &      0.266  \\
\vspace{0.1cm}  Cra2   &  51 & 59        &    $-$0.246     &      0.052     &     $-$0.227     &      0.026  \\
\vspace{0.1cm}  Gru1   &  4  & 2        &     $-$0.254   &       0.220    &      $-$0.532    &       0.288  \\
\vspace{0.1cm}  Hya2   &  5 & 6         &     $-$0.417   &       0.402    &       0.179    &       0.339  \\

 \tableline
\end{tabular}
\tablecomments{Our measured PMs. Column 1 lists the name, followed by the number of stars from the spectroscopic sample of \citealt{Fritz2018} that pass our cuts, the number of new members that we add, and the resulting PM values in $\mu_{\alpha \star}$, $\delta \mu_{\alpha \star}$, $\mu_{\delta}$, and $\delta \mu_{\delta}$.}
\label{tab:PMs}
\end{center}
\end{table*}

\begin{table*}[t]
\begin{center}
\caption{Predicted PMs and Radial Velocities.\label{predrv}}
\begin{tabular}{lcccccccc}
\tableline
 \\
\vspace{0.1cm} name  &   $\mu_{\alpha, \mathrm{pred}}$   &  $\mu_{\delta, \mathrm{pred}}$  &  RV$_{\mathrm{pred}}$ &   Nstars  &  $\mu_{\alpha, \mathrm{meas.}}$  &   $\mu_{\delta, \mathrm{meas.}}$ \\ 

\vspace{0.05cm}   &  (mas yr$^{-1}$) &  (mas yr$^{-1}$) &  (km s$^{-1}$) & &  (mas yr$^{-1}$) &  (mas yr$^{-1}$) \\

\tableline

\vspace{0.1cm} Pic2 & 1.69$_{-0.04}^{+0.05}$ & 0.77$_{-0.15}^{+0.13}$ & 333.83$_{-38.78}^{+42.84}$ & 2 & - & - \\ 
\vspace{0.1cm} Tuc4 & 1.26$_{-0.15}^{+0.17}$ & $-$2.19$_{-0.0}^{+0.02}$ & 33.51$_{-11.07}^{+14.66}$ & 0 & - & - \\ 
\vspace{0.1cm} Gru2 & 0.26$_{-0.16}^{+0.08}$ & $-$2.24$_{-0.02}^{+0.08}$ & $-$132.32$_{-2.23}^{+17.27}$ & 2 & - & - \\ 
\vspace{0.1cm} Tuc5 & 0.91$_{-0.14}^{+0.12}$ & $-$1.94$_{-0.02}^{+0.01}$ & 33.49$_{-10.64}^{+8.19}$ & 0 & - & - \\ 
\vspace{0.1cm} Sag2 & $-$0.01$_{-0.03}^{+0.03}$ & $-$1.3$_{-0.04}^{+0.02}$ & $-$339.54$_{-11.08}^{+6.35}$ & 0 & - & - \\ 
\vspace{0.1cm} Lae3 & 0.07$_{-0.06}^{+0.08}$ & $-$0.92$_{-0.01}^{+0.01}$ & $-$477.73$_{-13.26}^{+3.28}$ & 0 & - & - \\ 
\vspace{0.1cm} Hor2 & 1.25$_{-0.05}^{+0.06}$ & $-$0.65$_{-0.03}^{+0.03}$ & 159.81$_{-30.68}^{+26.76}$ & 0 & - & - \\ 
\vspace{0.1cm} Phx2 & 0.67$_{-0.08}^{+0.03}$ & $-$1.25$_{-0.01}^{+0.01}$ & $-$15.45$_{-10.48}^{+5.22}$ & 4 & $-$0.54 $\pm$ 0.10 & $-$1.17 $\pm$ 0.12 \\ 
\vspace{0.1cm} Eri3 & 1.04$_{-0.07}^{+0.06}$ & $-$0.75$_{-0.02}^{+0.0}$ & 126.18$_{-27.29}^{+29.89}$ & 1 & - & - \\ 
\vspace{0.1cm} Ret3 & 1.12$_{-0.05}^{+0.05}$ & $-$0.32$_{-0.04}^{+0.03}$ & 229.85$_{-32.02}^{+31.41}$ & 2 & - & - \\ 
\vspace{0.1cm} Ind1 & 0.26$_{-0.01}^{+0.01}$ & $-$0.98$_{-0.01}^{+0.01}$ & $-$93.93$_{-1.28}^{+1.88}$ & 4 & $-$0.23 $\pm$ 0.15 & $-$1.22 $\pm$ 0.15 \\ 
\vspace{0.1cm} Pic1 & 0.77$_{-0.02}^{+0.01}$ & $-$0.1$_{-0.01}^{+0.08}$ & 176.68$_{-20.77}^{+16.89}$ & 3 & $-$0.08 $\pm$ 0.24 & 0.07 $\pm$ 0.31 \\ 
\vspace{0.1cm} Col1 & 0.16$_{-0.0}^{+0.01}$ & $-$0.1$_{-0.0}^{+0.0}$ & 340.17$_{-13.5}^{+18.71}$ & 4 & $-$0.42 $\pm$ 0.14 & $-$0.15 $\pm$ 0.19 \\ 
\vspace{0.1cm} Peg3 & 0.12$_{-0.02}^{+0.02}$ & $-$0.27$_{-0.01}^{+0.0}$ & $-$327.12$_{-0.97}^{+8.9}$ & 2 & - & - \\

 \tableline
\end{tabular}
\tablecomments{Predicted PMs and radial velocities from the simulation for the set of dwarfs without measured radial velocities, and which don't have a probability of zero of being associated with the LMC based on their large distances from its orbital plane. The first column lists the name, followed by the PM and radial velocity predictions. The next three columns list the results of running our pipeline, using the predicted PM values as a starting point, on DR2. Column 5 lists the number of stars found to be consistent with this prediction, and if more than 3, then their weighted average PMs.}
\label{tab:pred}
\end{center}
\end{table*}

\section{Membership Selection and Proper Motion Determination} \label{sec:data}

We retrieved data within three times the half-light radii of each dwarf galaxy (see Table~\ref{tab:sample}) from the Gaia archive using \texttt{pygacs}\footnote{\small \url{https://github.com/Johannes-Sahlmann/pygacs}}. Following \citet{Helmi2018}, we first clean the source lists of stars with \texttt{visibility-periods-used} $< 5$ and relatively well-measured parallaxes, indicative of MW foreground stars, using $\omega - 2\sigma_{\omega} > 0$.



We next consider a set of nested criteria for our membership selection based on the publicly available spectroscopic member catalogs, the PMs, and the position in the color-magnitude diagram (CMD). We start with the sample of spectroscopic target lists compiled and presented in \cite{Fritz2018}. These lists contain stars that were deemed spectroscopic members as well as non-members by the original authors. We then pull stars from our DR2 source lists that match the weighted-average PMs of the spectroscopic members to within $2\sigma$ where $\sigma$ is the variance. 
PARSEC isochrones \citep{Bressan2012}, with ages and metallicities compiled from the literature, are then used to define a region in the CMD where member stars are expected to populate it. Only stars whose position in the CMD is compatible with the used isochrones are considered as possible members. We use $2\sigma$ as the maximum allowed distance from the star to the isochrone. We then perform these same steps on a ``background'' field, chosen to have the same galactic longitude as the dwarf under consideration, but opposite galactic latitude. Stars in the background field that pass the same PM and CMD cuts as the target field are used to construct a ``Hess'' diagram as follows: since we are dealing with quite sparse CMDs in general, we choose a fairly coarse grid of $5\times5$ bins. We then use the total number of background stars in the CMD and this number of bins to set an average density. This average density is assigned to any bin that contains zero stars. Then for each target star we interpolate over the 4 closest bins to assign an expected density at that point. We then use this density to weight the PM error of each star, i.e., multiply the PM errors by the ratio of the average to the expected density. 

Unsurprisingly, we pick up many potential member stars in fields closer to the disk. We therefore impose a strict additional cut that any new members must be within a radius of $1r_h$ of the dwarf center. This rejects many stars and cuts out the possibility of detecting any tidal signatures, but here we are aiming only for accurate center-of-mass PMs, and hence choose to focus on the cores of the dwarfs.  At the end of this procedure, we ensure that our newly-minted  members have not already been targeted and rejected as non members in the original spectroscopic samples. 

In summary, the method involves the following steps: (1) obtain clean source lists within $3r_h$ of each dwarf; (2) clean for parallax following \citet{Helmi2018}; (3) obtain an initial set of candidate member stars using the weighted average PM of the spectroscopic members plus or minus $2\sigma$ as the tolerance; (4) perform cuts keeping only stars that match the dwarf CMD to within $2\sigma$; (5) perform an identical set of steps on a background field; (6) weight PM errors of each target star by background field density at that location; (7) perform a cut in projected distance from the dwarf center of $1r_h$; (8) ensure that any new members that pass all these cuts were not already rejected in the original spectroscopic campaigns.

\begin{figure}
\begin{center}
\includegraphics[width=3.5in]{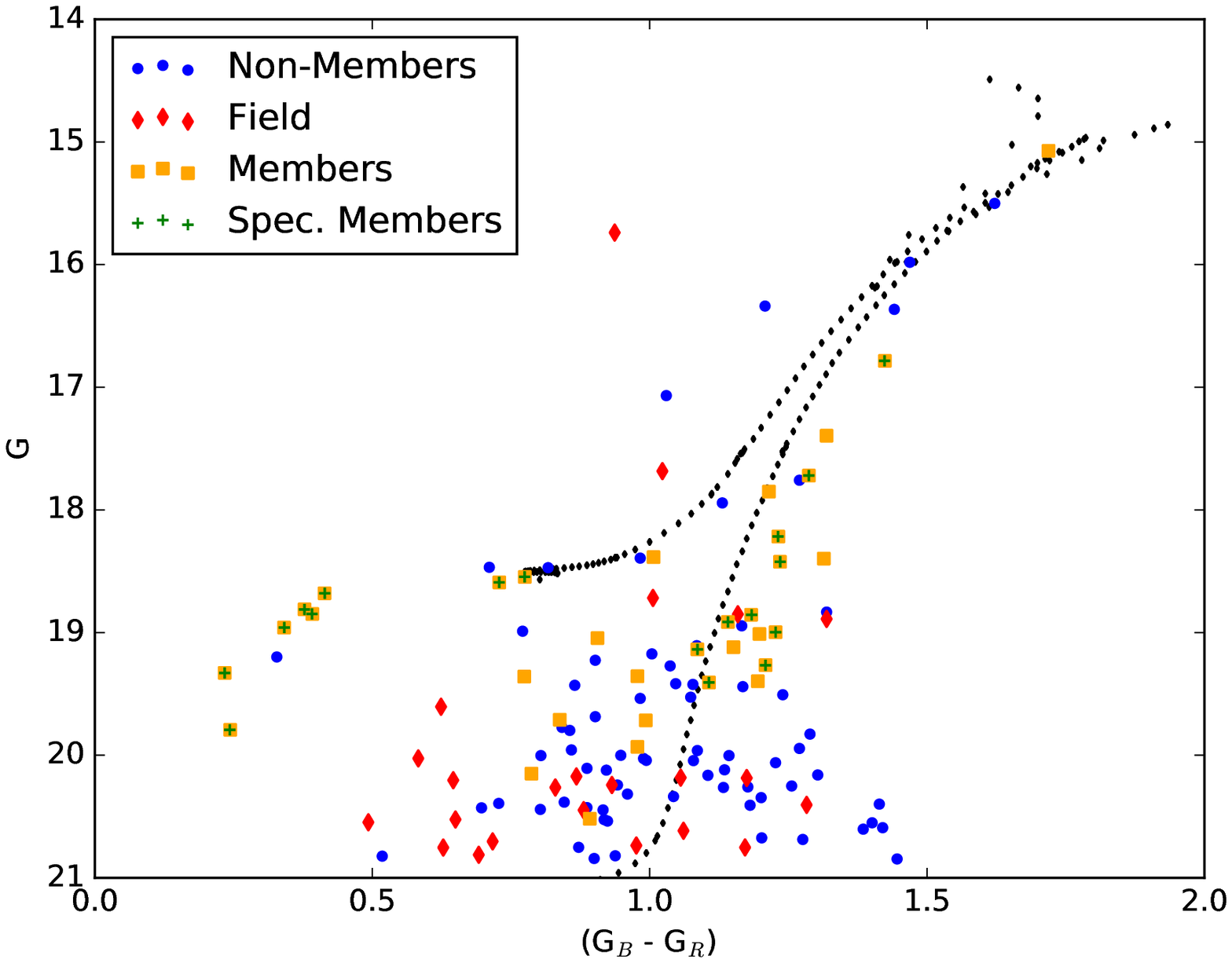}
\includegraphics[width=3.5in]{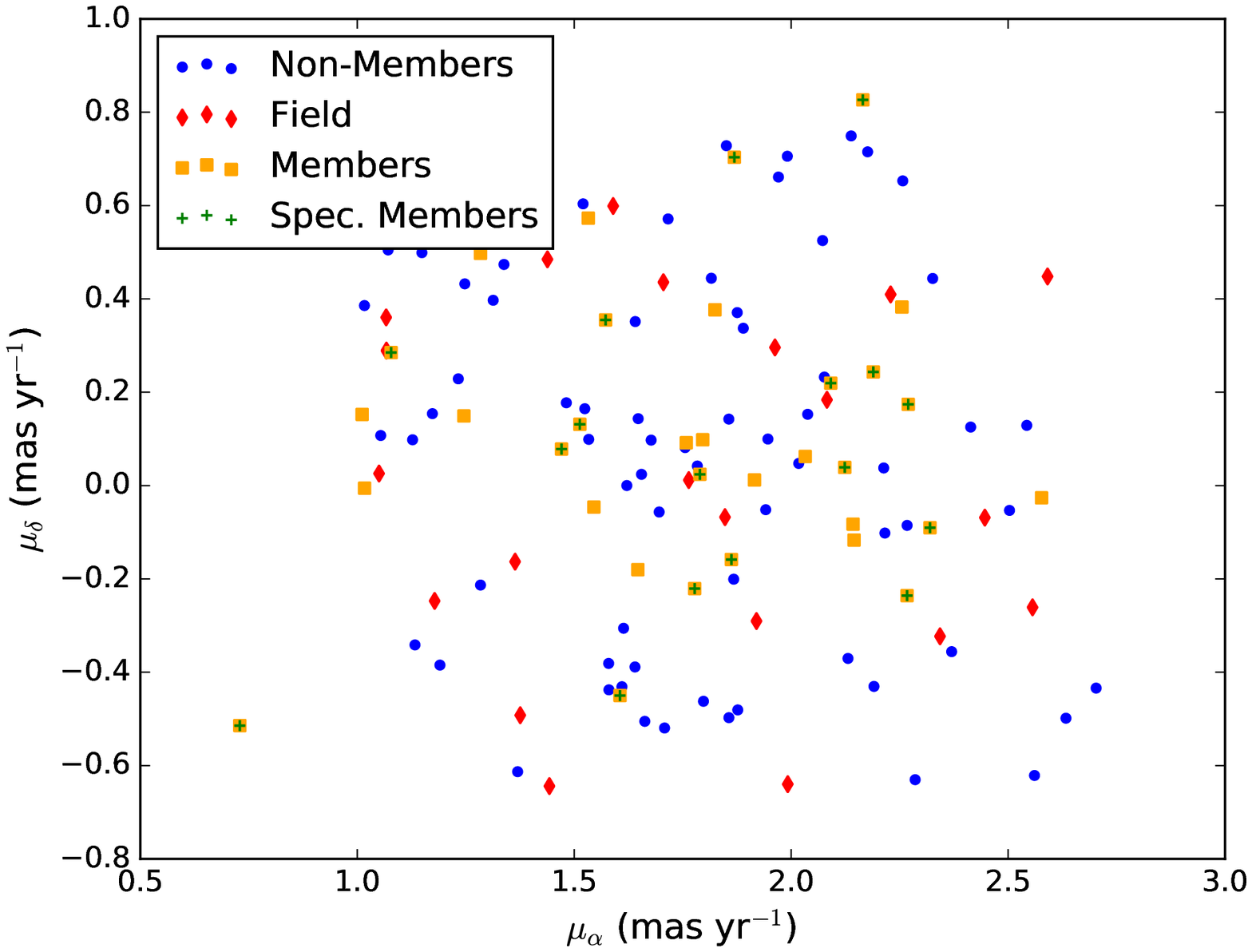}
\caption{Top: Our selection process for Car2. Green plus signs show the spectroscopic member stars that make it through our quality cuts, yellow, blue and red stars are consistent with the spectroscopic PMs. Red stars are from our control 'background' field, blue stars are target members that get cut due to our $1-r_h$ radius selection, and yellow stars (without plus signs overlaid) are 14 newly added members, used in addition to the spectroscopic members for the PM calculation. A PARSEC isochrone is overlaid for illustrative purposes. Bottom: The PM field, with the same color scheme.}
\label{fig:selection1}
\end{center}
\end{figure}

\begin{figure}
\begin{center}
\includegraphics[width=3.5in]{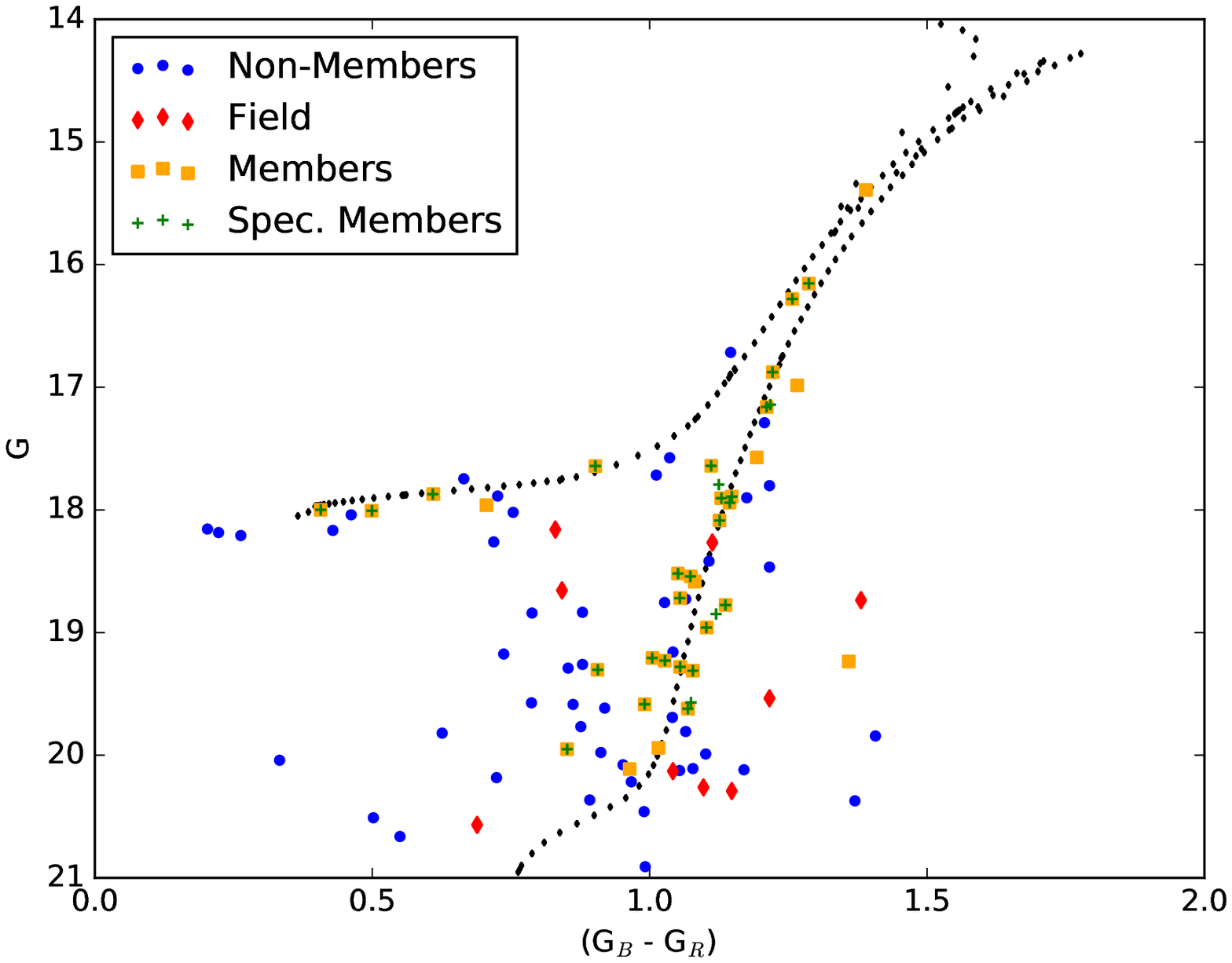}
\includegraphics[width=3.5in]{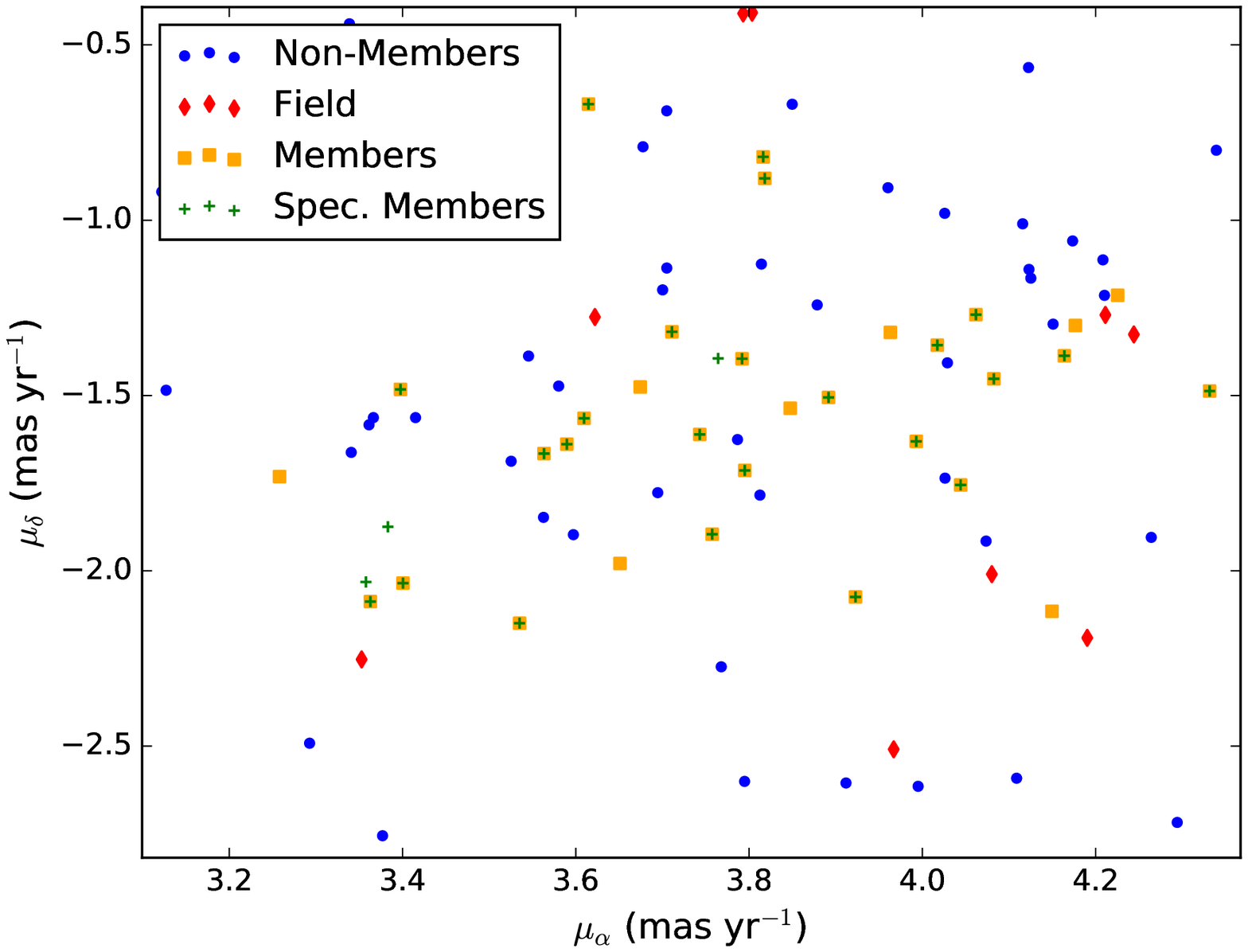}
\caption{Top: Same as Fig.~\ref{fig:selection1} but for Hyi1. In this case we add 8 new members to the 30 spectroscopic members, some of which are relatively bright. Note that 4 of the 30 spectroscopic members do not pass our parallax quality cuts and are therefore not used in our analysis, but we show them here (green plus signs) for completeness. 
}
\label{fig:selection2}
\end{center}
\end{figure}

\begin{figure}
\begin{center}
\includegraphics[width=3.5in]{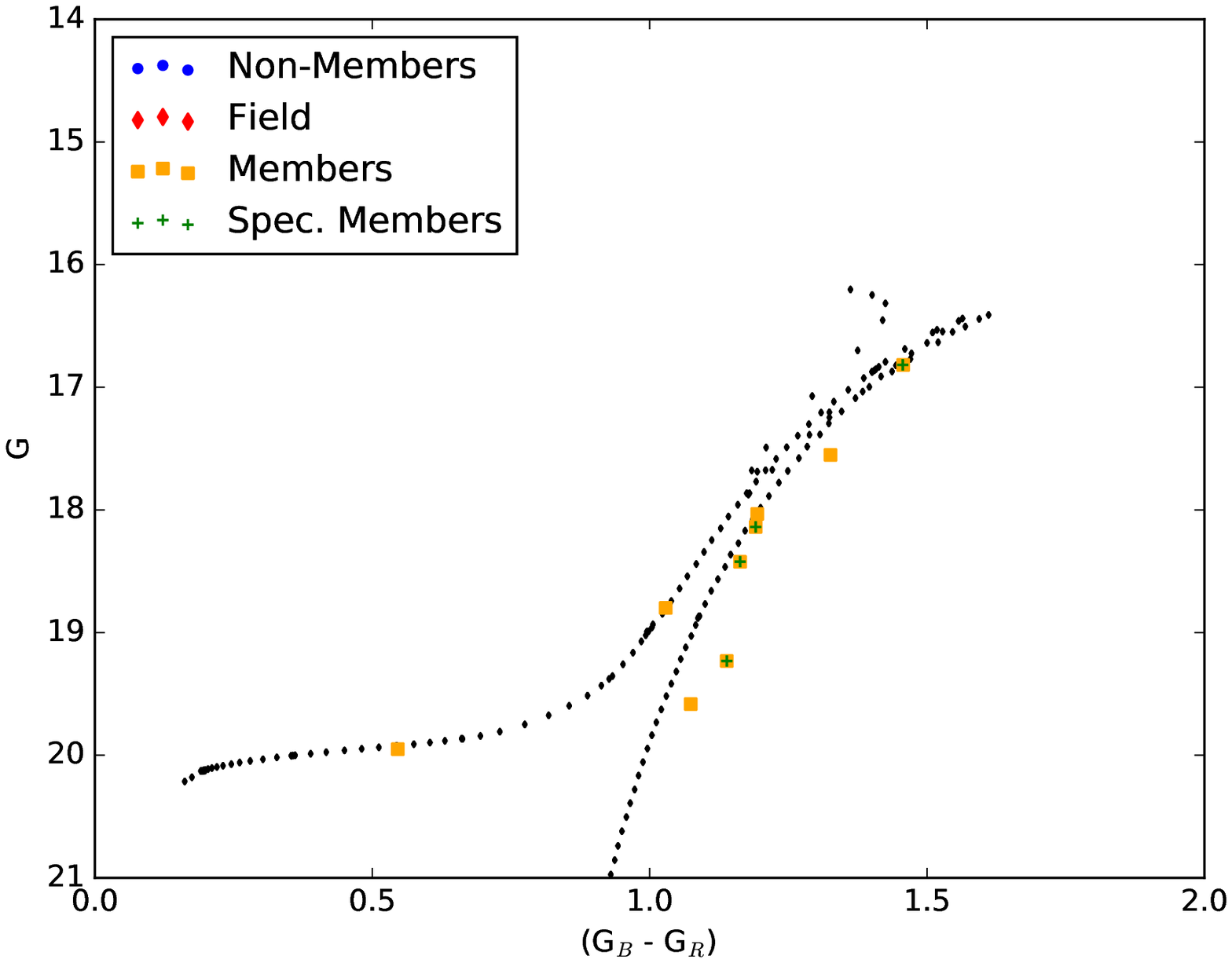}
\includegraphics[width=3.5in]{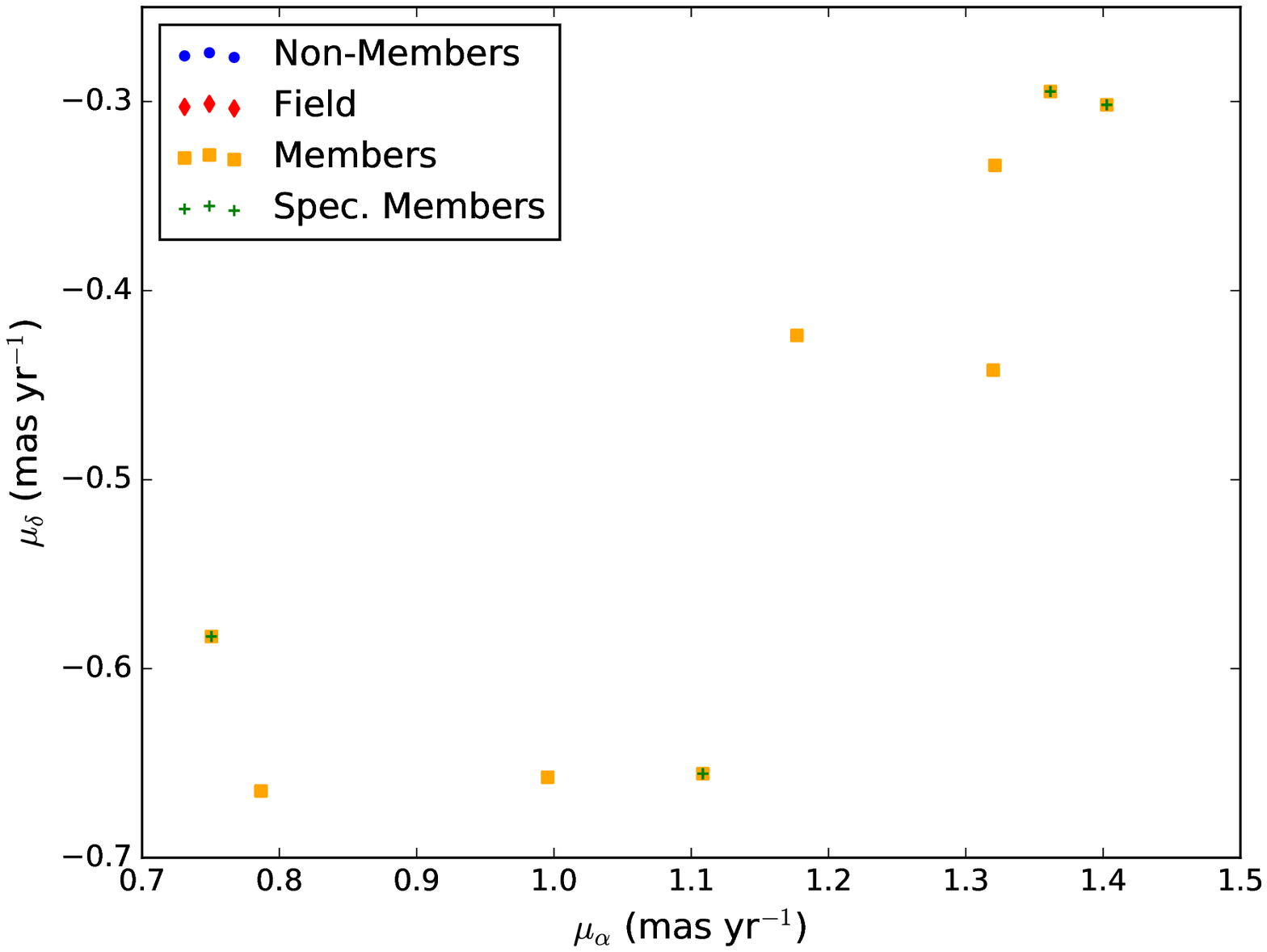}
\caption{Same as Figs.~\ref{fig:selection1} and \ref{fig:selection2} but for the more modest gains in the case of Hor1. We are however adding relatively bright (and therefore relatively well-measured) stars.}
\label{fig:selection3}
\end{center}
\end{figure}

In Figs.~\ref{fig:selection1}, \ref{fig:selection2} and \ref{fig:selection3} we show three illustrative examples of this process. In the figures, we show a CMD of the spectroscopic members (green plus signs), the new members (yellow squares), the control field stars (red diamonds) and the candidate members rejected by the projected distance cut (blue circles). In Fig.~\ref{fig:selection1} we show the case for Car2 in which we successfully add 14 new members to the spectroscopic sample thereby decreasing the final errors by a factor of 2. In Fig.~\ref{fig:selection2} we show the case for Hyi1, in which we add 8 new members to the spectroscopic sample of 30 from \cite{Koposov2018}. Four of these 8 new member stars are quite bright, and therefore constraining, as can be seen in the figure. Finally, in Fig.~\ref{fig:selection3} we show a third example of a dwarf (Hor1) where we don't add that many new members, but where the result is still interesting given how sparse the current sample is.


In the vast majority, we add only between 3 and 5 stars, but some of these are bright and therefore useful additions to these sparse samples, affording modest improvements of the PM errors. However, in many fields we add mostly faint stars, on the limit of what Gaia can do in DR2. A good example is Cra2, a field in which we add many stars (59), but the majority are very faint such that the resultant errors are barely an improvement over the spectroscopic-only sample PMs. 
For this field we impose a magnitude cut at $G>20$. For all other fields we do not employ this cut so as to recover the faint spectroscopic sample. 


Table~\ref{tab:PMs} presents our measurements, including how many stars are added for each field. Most values are consistent within the errors of the spectroscopic samples presented in \cite{Fritz2018} and \cite{Simon2018}. 
Exceptions to this consistency include our value for Tuc2, which differs at the $3\sigma$ level in $\delta$ from \cite{Fritz2018} and at the $2.5\sigma$ level also in $\delta$ from \cite{Simon2018}, but is very consistent in $\alpha$. 
We do consider and attempt to measure PMs for Peg3 and Eri2, both of which have measured RV's. However, for Peg3 \cite{Fritz2018} find no spectroscopic matches, and for Eri2, which is at 380 kpc, our pipeline to find additional members only produces 2 additional stars. The resulting PM errors are still in the $\sim 0.3 \masyr$ range, which at that distance corresponds to $\sim 540 \kms$, hopelessly large. This large error, in conjunction with the fact that membership can be ruled out via its location and distance alone (see Section~\ref{sec:results}), prompt us to eliminate it from further analysis.

Soon after submission of this work, \citet{Massari2018} presented a paper searching for additional photometric members for seven dwarfs in DR2, with two dwarfs overlapping with our sample, Car2 and Ret2. Their method is not dissimilar to ours. For these two dwarfs, they also choose additional members based on an initial RV guess, which is refined iteratively by $2.5\sigma$ cuts in PM and parallax, followed by a further culling on the basis of a color-magnitude diagram, and projected distance from the dwarf center. Our method also utilizes PM, CMD and projected distance cuts (though there are differences in choice of tolerance -- we choose $2\sigma$ for our PM and CMD cuts). A difference is that we do not further iterate on the selection, as we found that this mostly added faint stars, 
while they do not apply their selection on a control field, i.e., weight their results by a control field. Nonetheless, our results are very consistent with theirs. The reason that they report more new candidate members than we do is because of the very strict cut ($1-r_h$) that we make in projected distance from the dwarf center. Massari \& Helmi do not report what projected distance cuts they use, but if we do not make this cut then we obtain similar numbers of candidate members as they do: 47 total members for Ret2 and 61 for Car 2. The resulting errors from using this larger sample of candidate members are also consistent with their errors, and are slightly larger than those presented here with the $1r_h$ cut. This is consistent with our finding that adding many additional photometric members, that are mostly faint, do not improve the precision. This explains why we concentrated on searching for bright members very close to the UFD core. We do note, however, that the Massari \& Helmi method will be helpful to train further RV studies of these dwarfs and to better understand their structural properties.  

\citet{Helmi2018} have presented DR2 PMs for 9 classical dwarf spheroidals of the MW. Even though the regime, in terms of number of stars, is rather different than for the UFDs presented here, we mention their work here in order to provide a general consistency check of \textit{Gaia} PMs. Specifically, they are able to compare their results to previous studies using independent instruments/methods (both ground-based and with HST), and find that, in general, their \textit{Gaia} PMs are roughly consistent with previous determinations.

Galactocentric quantities are calculated using the same Cartesian coordinate system (X, Y, Z) as in \citet{NK13} \citep[see also][]{sohn12,sohn13}, and are mimicked when making predictions from the Aquarius simulation. In this system, the origin is at the Galactic center, the X-axis points in the direction from the Sun to the Galactic center, the Y-axis points in the direction of the Sun’s Galactic rotation, and the Z-axis points toward the Galactic north pole. The position and velocity of the dwarfs in this frame can be derived from the observed sky positions, distances, line-of-sight velocities, and PMs. Errors in the Galactocentric quantities are calculated by doing 100,000 Monte Carlo drawings over the errors in the measured PMs, radial velocities and distance moduli. Solar parameters are from \cite{bovy12}.

\section{Results} \label{sec:results}

\subsection{Simulation Set Up} \label{subsec:simulation_setup}

\begin{figure*}
\begin{center}
 \includegraphics[width=6.8in]{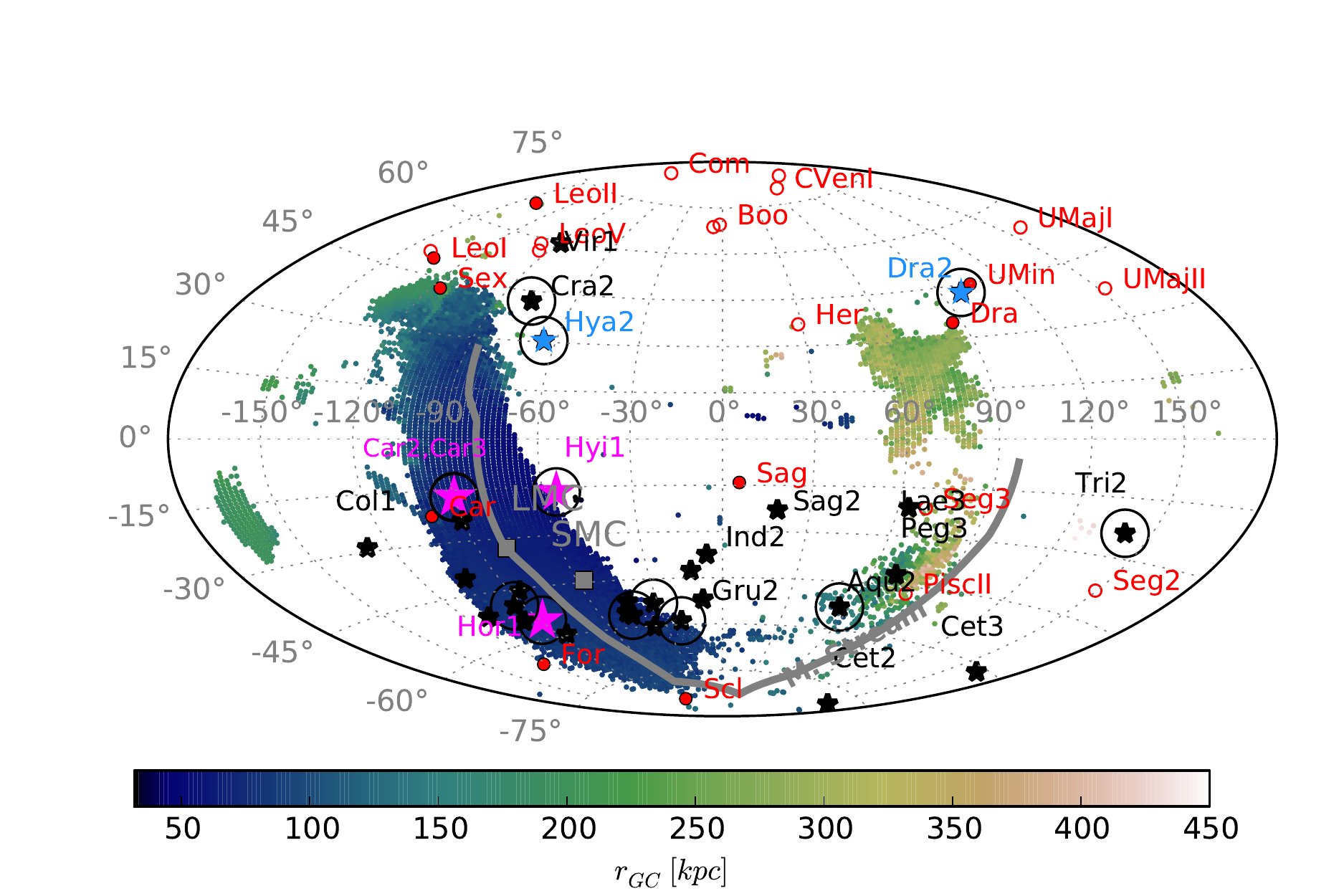} 
 \caption{The location of debris associated with LMCa prior to infall (colored points), now that the LMC is just past first pericenter, and color-coded by Galactocentric distance. Particles previously associated to the LMCa at infall outline today a clear stream on the sky that follows roughly the location of the Magellanic Stream (thick gray line) and with a well defined distance gradient (color bar). Over-plotted are the previously known satellites of the MW (red circles) with the ``classical'' dwarfs shown as filled-in red circles. The newly discovered dwarfs that are the subject of this work are shown with black stars and encircled black stars for those with a successful proper motion
 measurement. Dwarfs with kinematics consistent with an LMC association are highlighted in magenta (Hyi1, Car2, Car3 and Hor1) and in light blue we show Hya2 and Dra2 which membership certainty deserves more analysis in the future. See text and Figs.~\ref{fig:orbitalpole} and \ref{fig:vxyz} for more details.}
   \label{fig:aitoff}
\end{center}
\end{figure*}

We use the Aquarius simulations \citep{Springel2008} to identify LMC analogs within the cosmological scenario of $\Lambda$CDM. The Aquarius halos consist of six zoom-in DM-only cosmological simulations of MW-sized halos, with virial masses in the range $0.8 - 1.8 \times 10^{12} M_\odot$. The closest LMC analog, the LMCa, was chosen to match the present-day position and measured orbital velocity of the LMC \citep{NK06a}, and has a virial mass of $M_{200} = 3.6 \times 10^{10} M_\odot$, corresponding to a circular velocity of $65 \kms$. The corresponding host halo has a virial mass of $M_{200} =
1.8 \times 10^{12} M_\odot$ at $z = 0$, on the upper end of current
MW mass estimates \citep{blandhawthorn16}, but consistent with early analyses of Gaia DR2 data (\citealt{Watkins2018}, \citealt{Simon2018}, \citealt{Fritz2018}). Since this is a fully cosmological simulation, we know the full orbital history of LMCa as well as that of all DM particles that are bound to it pre-infall. The distance and velocity of LMCa most closely match those of the real LMC when it is at first pericenter (which in the simulation occurs at $t= 9.6$ Gyr), but is still consistent with measurements during a second pericenter passage at $t=13.3$ Gyr. We therefore consider mostly here the scenario of first infall, but discuss the implications for the case of a second passage as well. The MW analog has $M_{200} = 1.4$ and $1.6 \times 10^{12} M_\odot$ at the times of these first and second pericenter passages. More details on the identification of our LMCa can be found in S17 and \cite{sales11}.

We follow the evolution of LMCa using the level $3$ of the Aquarius halo A (Aq-A-3) with a mass and space resolution $4.9 \times 10^{4} M_\odot$ and $120$ pc respectively. Using the merger trees we trace backwards in time the LMCa orbital evolution and identify all particles that were initially bound to this structure before its infall onto the host halo. All particles associated to the LMCa before infall (defined as the last snapshot where the LMCa was the central of its own friends-of-friends group) provide a fair sampling of the phase space properties expected of any material initially associated to the LMC. In particular, since subhalos roughly follow the dark matter outside the first inner kpc of halos, looking at the distribution of the tagged particles at the time of first or second pericenter passage provide useful predictions for the present day positions and velocities to be expected of any dwarf companion that was brought onto the MW as part of the LMC group (see S17 for a more detailed discussion).  

Our approach relies on two assumptions: i) baryons will not statistically bias the orbital properties of LMC analogs in a sample of MW-like hosts, and ii) satellites of the LMC follow the same radial distribution as that of its dark matter halo before infall into the MW. Regarding the first point, we note that \citet{Patel2017} have shown that LMC analogs in Illustris 1 (with baryons) versus with Illustris Dark 1 (only dark matter) do not exhibit markedly different orbital properties, providing further validation to our analysis. The second point deserves more caution as baryonic processes have been shown in some cases to preferentially disrupt inner satellites due to the increased tides associated with the disks (e.g., \citealt{Donghia2010}, \citealt{Garrison-Kimmel2017}, \citealt{Ahmed2017}). Encouragingly, these effects are presumably smaller for a less massive disk such as the LMC and are increasingly weaker with distance to the host.  Although this is not well known in dwarf galaxies, for more massive systems the radial distribution of {\it luminous} satellites is expected to show little systematic difference with the dark matter beyond $\sim 10$\% of the virial radius of the host \citep{Sales2007, Vogelsberger2014}.  Assuming a virial radius of $130$ kpc for the LMC before infall, it would limit the applicability of our results for dwarfs closer to the LMC than $13$ kpc (see S17). All dwarfs explored in this paper, and in particular those considered likely members of the LMC group, lie well beyond this limit, with the closest being Car2 at $\sim 25$ kpc.

\subsection{Comparison to LMC Debris} \label{subsec:lmcdebris}

At the time of first pericenter passage for the LMCa, some tidal disruption due to the host halo has already set in and has started partially unbinding the group. These unbound particles however, follow a very well defined pattern on the sky, distance and velocity space due to the common orbital properties with the LMCa that can be compared to observations to determine which of the new dwarfs match these predictions. Fig.~\ref{fig:aitoff} shows the footprint of the LMCa debris on the sky at the time of first pericenter. The DM particles lie along a well-defined tidal tail which is roughly coincident with the real Magellanic Stream, sketched in with a thick grey line \citep{nidever10}. Over-plotted are the Galactic coordinates of the 32 new candidate dwarfs (star symbols). The spatial distribution of the new dwarfs coincides quite well with the sky distribution of DM particles that are (or were) bound to the LMC. 

The DM particles are color-coded by Galactocentric distance. There is a clear gradient in distance along the Stream. LMC debris can be distributed as far out as 300 kpc even at first pericenter. Nonetheless, because LMC debris still must lie close to the orbital plane of the LMC, Fig.~\ref{fig:aitoff} shows that sky distribution and distance by themselves are quite good determinants of LMC membership. 
We find that no particles in our LMC-analog cover the region of sky inhabited by Eri2, Ind2, Cet2, Tri2, Cet3 and Vir1. In these cases the probability of association is formally zero in the simulation, and the LMCa stream does not predict any dwarf/particle in those regions. 

\begin{figure}
\begin{center}
\includegraphics[width=3.5in]{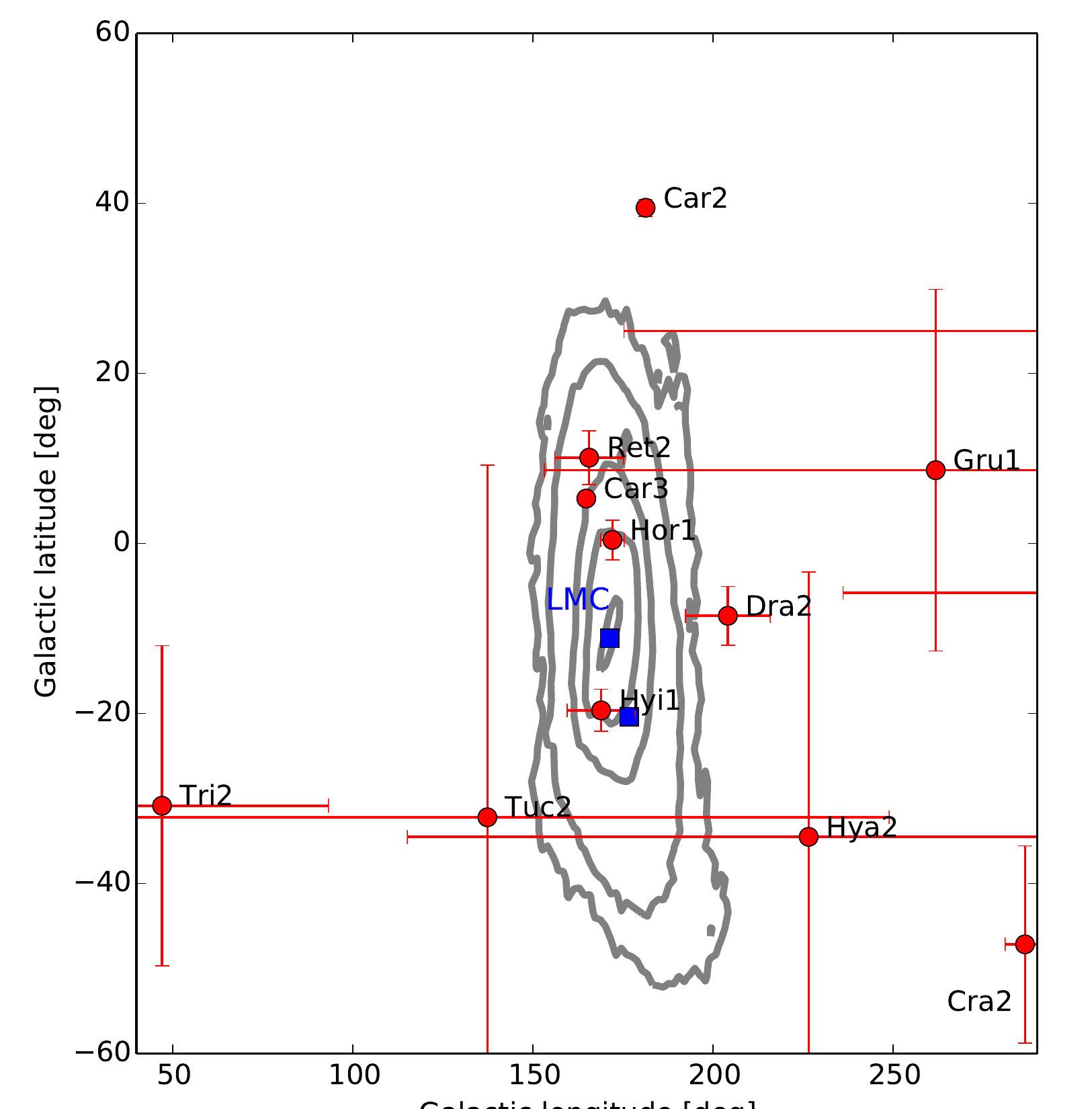}
\caption{The direction of the axis of orbital angular momentum of LMCa debris (grey density contours enclose from innermost to outermost lines 5 to 95 per cent of
all LMCa particles, respectively) along with those measured for dwarfs in this study that lie close to this direction. Hor1, Hyi1, Car3, and Ret2 are clearly coincident with the Magellanic system, a condition necessary but not sufficient for a common origin with the LMC. Dra2, Tuc2, Hya2 and Gru1 are marginally consistent given their measurement errors.}
\label{fig:orbitalpole}
\end{center}
\end{figure}

Certainty of membership for the remaining dwarfs, however, comes from adding the measured 3D velocities (radial velocities plus PMs) to the sky positions and distances, to get full 6-D quantities, and from comparing these 6-D quantities to the simulation predictions. We first compare the orbital poles of the satellites with 3D measurements to the orbital pole of the LMCa system. Fig.~\ref{fig:orbitalpole} plots the galactic $l$ and $b$ position of the orbital poles, which are preserved and should be consistent for in-falling groups, for LMCa particles (grey density contours) and for the subset of measured dwarfs that also inhabit this region of angular momentum space. Hor1, Hyi1, Car3, and Ret2 are clearly consistent with the angular momentum direction of the LMC system, listed in descending order of significance as represented by the density of LMCa debris in the same region. Dra2, Tuc2, Hya2 and Gru1 are also consistent within their measured $1\sigma$ errors, where the errors in the poles are calculated as standard deviation of 1000 Monte Carlo drawings over the measurement errors in Galactocentric $X$, $Y$, $Z$ and $V_X$, $V_Y$ and $V_Z$. 

\begin{figure*}
\begin{center}
\includegraphics[width=6.5in]{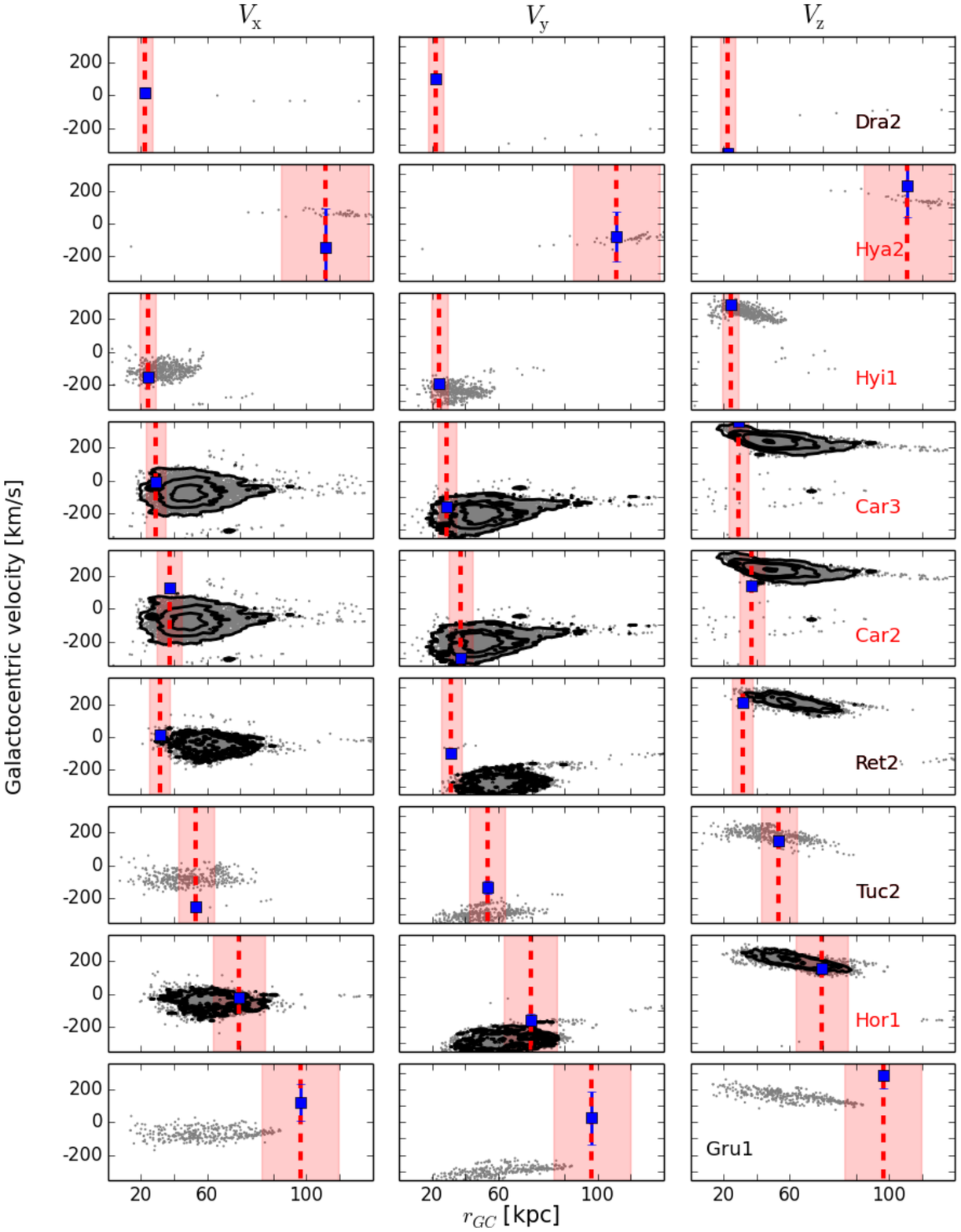}
\caption{Galactocentric velocity in $V_X$, $V_Y$ and $V_Z$ ($\kms$) versus Galactocentric distance (kpc) for the observed dwarfs versus LMCa debris from the simulation, for 9 dwarfs of interest (labeled in each right-hand side panel) sorted by galactic latitude. Gray dots represent the LMCa dark matter particles located within $5$ deg$^2$ of the position of the dwarf. Red dashed lines correspond to the observed radial distance of each dwarf together with the $\pm 20\%$ range used to average the predictions in Table~3. The observed velocities are indicated with blue square symbols. Hyi1, Car3, Car2 and Hor1 are likely associated to the LMC (red labels) whereas Gru1, Tuc2 and Ret2 are currently disfavored. Hya2 and Dra2 cannot be ruled out and deserved further analysis (see text for more detail).}
\label{fig:vxyz}
\end{center}
\end{figure*}

We next consider the Galactocentric radial and tangential velocities of the debris, and compare to those measured for the dwarfs. Following S17, we select all DM particles once part of the LMCa system that lie within a $5^{\circ}$ radius of the sky positions of the dwarfs. We check their Galactocentric distances, and their Galactocentric $V_X$, $V_Y$ and $V_Z$ values. We then plot, in Fig.~\ref{fig:vxyz}, the corresponding observed distance of the dwarf $\pm 20\%$ range (red regions) and the observed velocities and errors. 
The radius and distance tolerance were chosen to maximize the total number of potential LMCa particles in a given region of sky while still giving peaked (confined) predicted velocity histograms.

At least 4 dwarfs show clearly consistent positions and velocity measurements compared to our predictions: Hyi1, Car3, Car2 and Hor1 (red labels), presenting a compelling case of probable membership to the LMC group. Ret2, Tuc2 and Gru1 have velocity components that are not consistent within $3$ sigma of our predictions and are unlikely members of the LMC group according to our analysis. Notice that Hya2 is an interesting case. In a first pericenter passage scenario, Hya2 occupies the foremost tip of the leading arm of the stream. Only a few particles are expected in that area of the sky (see Fig.~\ref{fig:aitoff}), but despite this seemingly low chance of association, their velocities agree well with the observed ones for Hya2, suggesting that association of Hya2 to the LMC might not be quickly ruled out. On the other hand, Dra2 although consistent within 1 sigma with the position of the orbital pole (see Fig.~\ref{fig:orbitalpole}), has a location on the sky that is disfavored and we find almost no particles associated with the stream (however see discussion in Sec.~\ref{ssec:2ndper}). For completeness, we include this case in the first row of Fig.~\ref{fig:vxyz} for easy comparison to the other cases. 

\subsection{Association with the LMC in the case of second pericenter passage}
\label{ssec:2ndper}

A similar analysis as presented in the previous section can be done for the case of LMCa transiting through its second pericenter passage. We have checked that our results presented in Figs.~\ref{fig:aitoff} through \ref{fig:vxyz} are still valid even if a second approach is considered. Moreover, because of the more extended footprint of the stream over the North Hemisphere due to the completion of one full orbit for the LMC (see Fig. 1 in \citealt{sales11}), the chances of association of Hya2 are significantly improved. 

Worth mentioning, the dispersion of the stream on the sky during the second passage also allows a better consideration of the case for Dra2. We find that for that region of the sky, the LMCa debris predicts velocities consistent with those measured for Dra2. This might imply that had the LMCa been more massive (and therefore extended) the sky position of Dra2 could have been more sampled even in first pericenter passage. This does not remove the fact that such large separation from the main stream is less likely to represent a previous association to the LMC. Based on our results, we cannot rule out the association of Dra2, an issue that deserves further examination in the future.

\subsection{Predictions for galaxies without radial velocity measurements} \label{subsec:pred}

\begin{figure}
\begin{center}
\includegraphics[width=3.5in]{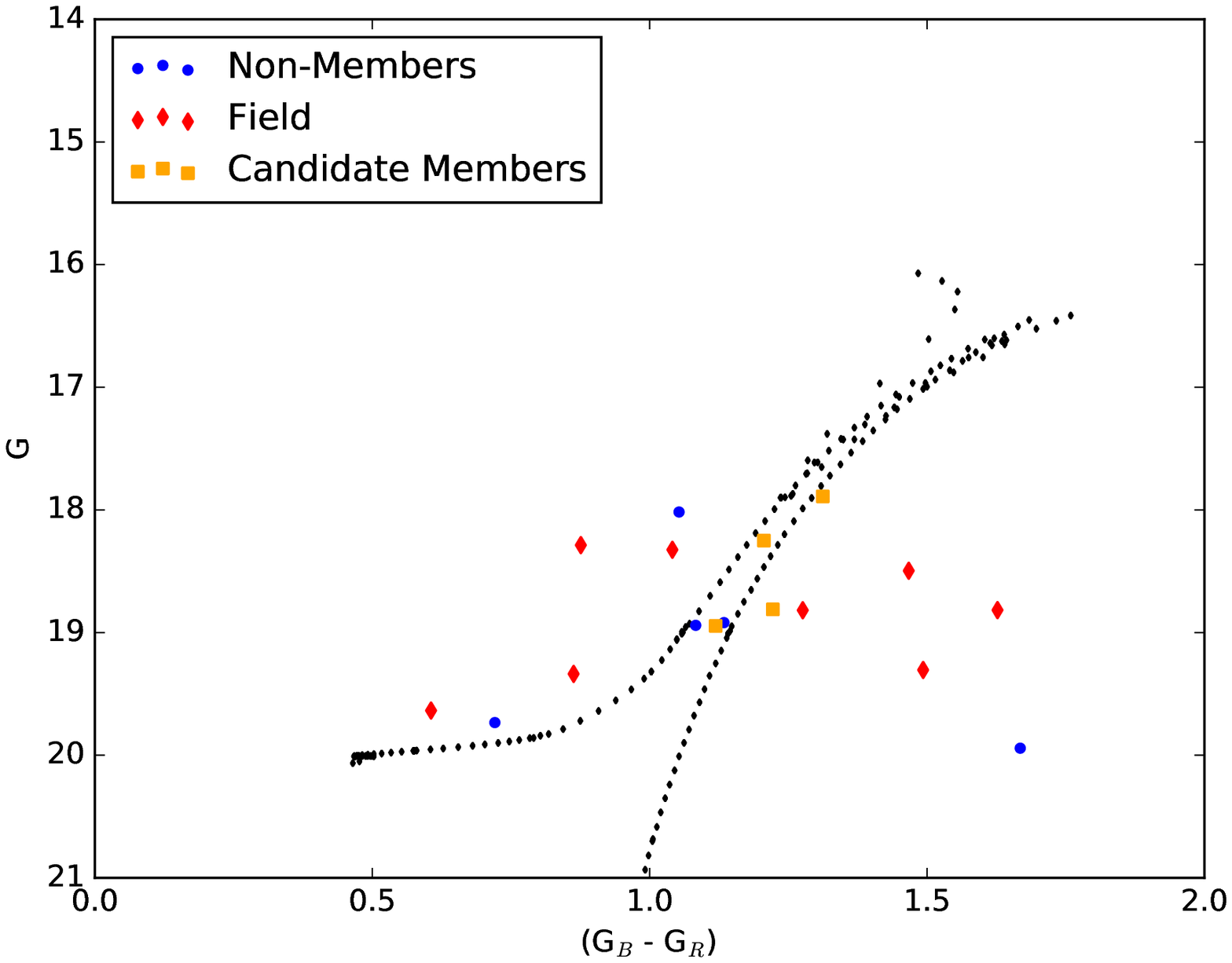}
\includegraphics[width=3.5in]{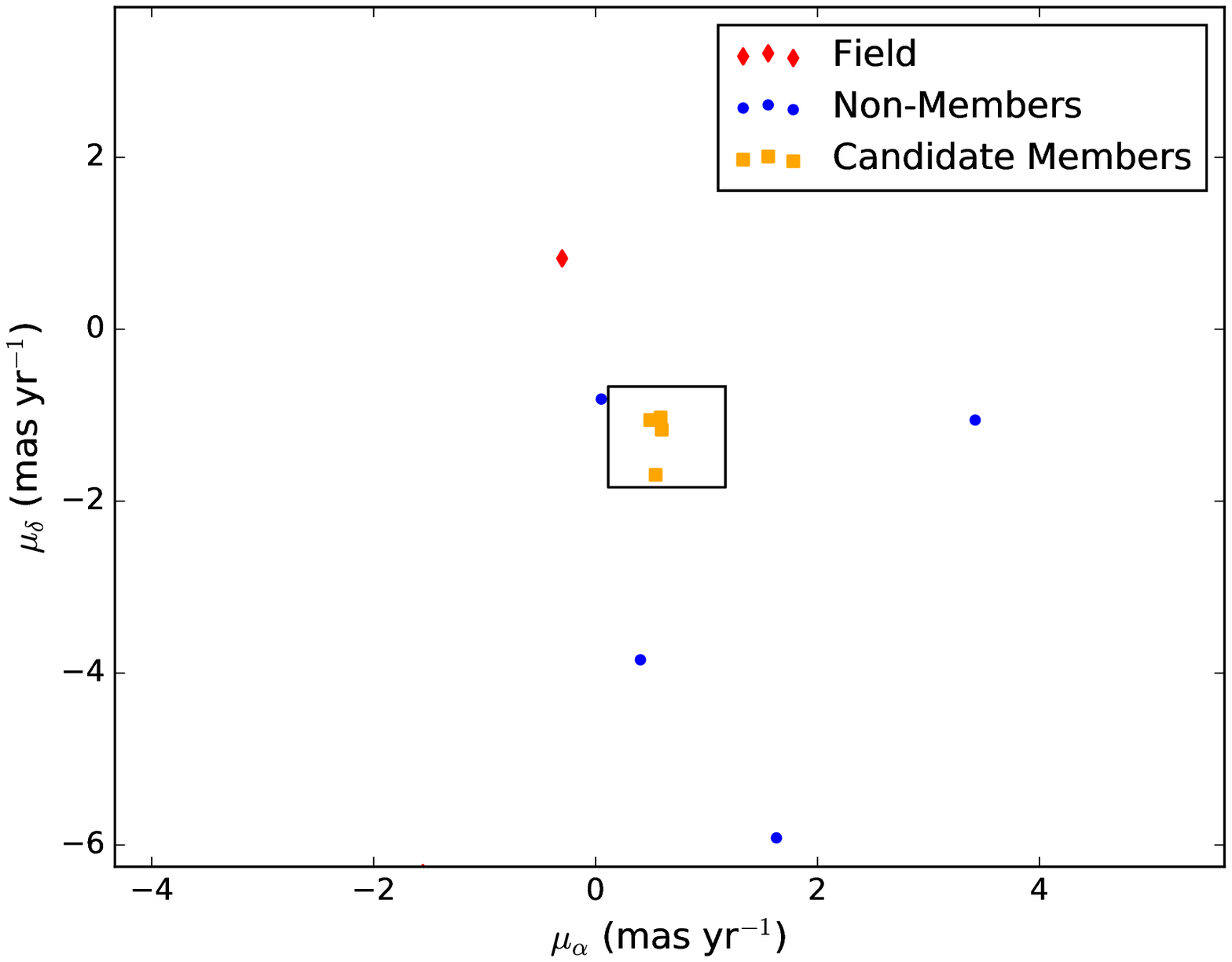}
\caption{Same as Figs.~\ref{fig:selection1} and \ref{fig:selection2} but for the prediction of Phx2. Blue points are candidate member stars, while red points show our control (background) field.}
\label{fig:Phx2}
\end{center}
\end{figure}

We now use the velocity information from the simulation to ascertain whether any of the galaxies in Table~\ref{tab:sample} without RV measurements might be associated with the Clouds. We convert the Galactocentric velocities measured in the simulation, along with the 25\% and 75\% bounds, for each of these dwarfs to observed parameters ($\mu_{\alpha \star}$, $\mu_{\delta}$ and radial velocity). Using these predicted PMs and a generous window around them as the starting point  (in lieu  of the PM of the spectroscopic sample), we then run our exact same pipeline as in Section~\ref{sec:data}. If we find more than 3 stars that pass all our cuts, we report their ``measured'' PMs in Table~\ref{tab:pred}. For the case of Phx2 we find a pretty convincing clump of stars that are consistent with its PM prediction, see Figure~\ref{fig:Phx2}. This would be an interesting target for radial velocity follow-up.

\section{Conclusions} \label{sec:conc}
We use an LMC analog in the Aquarius simulation to test whether any of the 32 newly discovered ultra-faint dwarf galaxies and candidate dwarf galaxies have 3D velocities, positions and distances consistent with having come in with that system. The missing piece of the puzzle were the proper motions, which we measure here for 13 UFDs using Gaia DR2.

Starting with the PM of the spectroscopic members, we attempt to identify additional member stars using a series of nested PM and CMD-space cuts. This approach works especially well for Hyi1 and Car2, with more modest gains for other systems.

Using the resulting 6D velocities, we find that 4 of these galaxies (Hor1, Car2, Car3 and Hyi1) have come in with the LMC. Ret2, Gru1 and Tuc2 do not match LMC debris within 3$\sigma$ in all their velocity components and so are not as favorable. We rule out Tuc3, Cra2, Tri2 and Aqu2 as potential members. 

Hya2 and Dra2 are interesting cases. Their orbital poles match well with the predictions from the simulation. In the case of Hya2, it occupies the foremost tip of the leading arm of the stream, where only a few particles are expected, but their velocities agree well with those observed. In the case of Dra2, we find almost no particles at its location in the simulation. However, had LMCa been more massive and extended, this area could be more populated with LMC debris even on a first passage, and therefore we cannot rule out the association of Dra2.

Of the dwarfs without measured PMs, 5 are deemed unlikely on the basis of their positions and distances alone (Eri2, Ind2, Cet2, Cet3 and Vir1). For the remaining sample, we use the simulation to predict proper motions and radial velocities, finding that Phx2 has an overdensity of stars in DR2 consistent with this PM prediction. It would be a good candidate for radial velocity follow up.


Since our LMCa is on the low side of the LMC mass, and our MW is on the high side, our conclusions are conservative in terms of association. The most promising case deserving further evaluation is Dra2 since a more extended LMCa could provide a chance of association at its slightly large angular separation on the sky.

The finding of 4 confirmed companions of the LMC, with perhaps 2 more, given the uncertainty in LMC mass, is consistent with the numbers expected for comparable mass systems in LCDM theory \cite{Sales2013}.

Recent work from the \textit{Gaia} Collaboration \citep{Helmi2018} shows that some of the classical dwarf spheroidals have PMs consistent with being accreted in a group. The analysis of \citealt{sales11} concluded that none of the classical dwarf spheroidals in the MW would be associated with the Clouds if on a first or second passage. However, with the new PMs, a reanalysis is warranted. An analysis of any new data on the bright classical dwarfs will be presented in a separate work.

\acknowledgements
We thank the anonymous referee for constructive comments that helped improve the presentation of the results. NK is supported by the NSF CAREER award 1455260. LVS acknowledges support from the Hellman Foundation. This work has made use of data from the European Space Agency (ESA) mission {\it Gaia} (\url{https://www.cosmos.esa.int/gaia}), processed by the {\it Gaia} Data Processing and Analysis Consortium (DPAC, \url{https://www.cosmos.esa.int/web/gaia/dpac/consortium}). Funding
for the DPAC has been provided by national institutions, in particular the institutions participating in the {\it Gaia} Multilateral Agreement. This project is part of the HSTPROMO (High-resolution Space Telescope PROper MOtion) Collaboration\footnote{http://www.stsci.edu/$\sim$marel/hstpromo.html}, a set of projects aimed at improving our dynamical understanding of stars, clusters and galaxies in the nearby Universe through measurement and interpretation of proper motions from HST, Gaia, and other space observatories. We thank the collaboration members for the sharing of their ideas and software.

\bibliography{Kallivayalil_bib}
\end{document}